\documentclass[conference,dvipdfmx]{IEEEtran}

\usepackage{graphicx}
\usepackage{cite}
\usepackage{bm,bbm}
\usepackage{algorithm,algpseudocode}
\usepackage{booktabs}
\def\ind{\mathop{\mathbbm{1}}\nolimits}

\usepackage{url}

\usepackage{amsmath,amssymb,amsfonts,mathtools,amsthm}

\DeclareMathOperator*{\argmax}{arg\,max}
\usepackage{comment}

\theoremstyle{definition}
\theoremstyle{definition}

\ifCLASSOPTIONcompsoc
	\usepackage[font=normalsize,labelfont=sf,textfont=sf]{subfig}
\else
	\usepackage[font=footnotesize]{subfig}
\fi

\begin{document}
\title{Deep Reinforcement Learning-Based\\Channel Allocation for Wireless LANs\\with Graph Convolutional Networks}

\author{
\IEEEauthorblockN{
\normalsize Kota Nakashima\IEEEauthorrefmark{1},
\normalsize Shotaro Kamiya\IEEEauthorrefmark{1},
\normalsize Kazuki Ohtsu\IEEEauthorrefmark{1},
\normalsize Koji Yamamoto\IEEEauthorrefmark{1},
\normalsize Takayuki Nishio\IEEEauthorrefmark{1}, and
\normalsize Masahiro Morikura\IEEEauthorrefmark{1}
}
\IEEEauthorblockA{
\IEEEauthorrefmark{1}\small Graduate School of Informatics, Kyoto University,
Yoshida-honmachi, Sakyo-ku, Kyoto 606-8501, Japan
}
}

\maketitle
\begin{abstract}
  Last year, IEEE 802.11 Extremely High Throughput Study Group (EHT Study Group) was established to initiate discussions on new IEEE 802.11 features.
  Coordinated control methods of the access points (APs) in the wireless local area networks (WLANs) are discussed in EHT Study Group.
  The present study proposes a deep reinforcement learning-based channel allocation scheme using graph convolutional networks (GCNs).
  As a deep reinforcement learning method, we use a well-known method double deep Q-network.
  In densely deployed WLANs, the number of the available topologies of APs is extremely high, and thus we extract the features of the topological structures based on GCNs.
  We apply GCNs to a contention graph where APs within their carrier sensing ranges are connected to extract the features of carrier sensing relationships.
  Additionally, to improve the learning speed especially in an early stage of learning, we employ a game theory-based method to collect the training data independently of the neural network model.
	The simulation results indicate that the proposed method can appropriately control the channels when compared to extant methods.
\end{abstract}
\IEEEpeerreviewmaketitle

%
%
\section{Introduction}
IEEE 802.11 Extremely High Throughput Study Group (EHT Study Group) was established to initiate discussion on new IEEE 802.11 features for bands between 1 and 7.125\,GHz\cite{EHT}.
In order to mitigate problems in existing wireless local area networks (WLANs) such as densely deployment problem, the coordinated control methods of the access points (APs) in the WLANs are discussed in EHT Study Group.

Channel allocation is an important problem in densely deployed WLANs since there are huge number of APs but the available channels are limited.
Bad channel allocation causes substantial contention among APs and stations (STAs) and reduces the throughput of each AP.
It is possible to avoid the problem by effectively allocating the limited channels to each AP.
This is the motivation for our study.

Since the throughput can be obtained only from the observations, we cannot model the objective functions.
Therefore, optimization approaches cannot be applied to this channel allocation problem.
As throughput-independent approaches, game theory-based approaches are proposed \cite{pg1, pg2} and are based on spatial adaptive play (SAP)\cite{sap} that determines the action to reach the Nash equilibrium wherein the payoff function is the highest.
However, game theory-based approaches cannot always ensure that the throughput is the highest because the control results depend on the validity of the model and not on the payoff function (which corresponds to the maximization target of game theory-based approaches).
We use reinforcement learning to allocate the channels based on the feedback of the measured throughput.
However, a reinforcement learning-based approach without transfer learning takes a long time to obtain a desirable policy if the agent selects actions based on the value function-based methods, (for e.g., $\epsilon$-greedy \cite{sutton}) because the value function is randomly initialized in the beginning.
We employ a few techniques, detailed subsequently, for WLANs channel allocation problem.

The contribution of the study involves proposing a deep reinforcement learning-based scheme that is suitable for channel allocation problems in high density WLANs.
As a deep reinforcement learning method, we use a well-known method double deep Q-network (DDQN)\cite{ddqn2}.
To extract the features of the adjacency of the APs and used channels, we introduce graph convolutional networks (GCNs) \cite{gcn3} as neural network layers.
Additionally, we use SAP to collect the training data to improve the learning speed.
The state should be adequately associated to the reward because the control of the reinforcement learning depends on the observed state.
We consider the graph as a state by considering APs within the carrier sensing range to adjoining, APs as nodes, and the adjoined APs as connected by edges.
This is based on an idea that the carrier sensing relationship between APs significantly affects the throughput.


%
%
\section{System Model}
\label{sec:system_model}

\subsection{Channel Allocation Problem in Wireless LANs}
\label{sec:csma}
Assume that $N$ APs are placed in a square-shaped region, and $M$ channels are available.
Let the index set of APs be denoted by $\mathcal{N} = \{1,2,\dots, N\}$, and the index set of available channels by $\mathcal{M} = \{1,2,\dots,M\}$.
It should be noted that $c_i$ corresponds to the one-hot vector of $M$ dimension, (e.g., if AP $i \in \mathcal{N}$ uses channel 2 $\in \mathcal{M}$, then $c_i = \left[ 0, 1, 0, \dots, 0 \right]^\mathsf{T}$).
In the system, a central controller is considered and is responsible for information gathering and channel allocation to each AP.
More specifically, the central controller observes and gathers the communication quality (e.g., the throughput), the adjacency relations of the APs, and the channel set $\bm{C} = [c_1\ c_2\ \dots\ c_N]$.
The study discusses the problem that the central controller successively searches the channel set that increases the communication quality.
The throughput of each AP decreases when the number of the APs using same channel within the carrier sensing range increases because of the carrier sense multiple access with collision avoidance mechanism\cite{csma}.
Thus, it is expected to increase the throughput by selecting the channels of the APs such that it decreases the number of the APs using the same channel within the carrier sensing range.

\subsection{Graph Structure of State}
\label{sec:graph}
We express the topology of APs, which consists of the APs and carrier sensing relationships, through a graph depicting $\mathcal{G} = (\mathcal{N}, \mathcal{E})$.
The edges $e_{ij} = \{i, j\} \in \mathcal{E}$ of the graph are connected if and only if the APs $i$ and $j$ are within the carrier sensing range.
We denote an adjacency matrix as an $N \times N$ matrix $\bm{A} = (A_{ij})$ as follows:
\begin{align}
  \label{eq:adj}
  A_{ij} = \begin{cases}
    1 & \text{if}\ i \neq j \land e_{ij} \in \mathcal{E}, \\
    0 & \text{otherwise}.
  \end{cases}
\end{align}

%
%

\section{Markov Decision Process}
\label{sec:MDP}
We define an MDP prior to the formulation of the reinforcement learning problem.
An MDP is defined as a quadruplet $\mathcal{(S, A, P, R)}$: $\mathcal{S}$ corresponds to the state space (which denotes a set of states in the environment); $\mathcal{A}$ corresponds to the action space (which denotes a set of actions adopted by the agent in each step); $\mathcal{P}$ corresponds to the transition probability to the next state given the current state and action; and $\mathcal{R}$ corresponds to the reward (which denotes a function of the current state, action, and the following state).
The agent determines the action based on the observed state in each discrete time step $t \in \mathbb{N}$.
Subsequently, it is desirable for the agent to act to transfer to a state that provides a high reward to the agent.
Therefore, the state should be adequately associated to the reward, and the design of the state corresponds to the vital point of an MDP.

Given that the throughput is significantly affected through the carrier sensing relationships between the APs, we design the state of the MDP based on the adjacency matrix $\bm{A}$ and $M \times N$ channel matrix $\bm{C} = [c_1\ c_2\ \dots\ c_N]$.
To reduce the number of available states, we determine isomorphism between graphs by comparing their canonical labels, which are detailed in Section~\ref{sec:map}.
Moreover, we design the action $a_t \in \mathcal{A}$ based on the index of the AP (which changes its channel) and the revised channel.
We consider the mean value of the throughput of a few APs in the ascending order of their throughput as a reward.

\subsection{State Mapping Method}
\label{sec:map}
To reduce the computational complexity, we reduce the number of states based on the canonical labeling \cite{iso, canonical}.
A graph can be represented in several different ways.
However, the canonical labels are identical if the graphs exhibit an identical topological structure and identical labeling of nodes and edges.
Thus, by comparing the canonical labels, we sort the graphs in a unique and deterministic way and consider two graphs as isomorphic if their canonical labels are identical.
With respect to the computation of automorphism and canonical labeling of graphs, we use an open source tool \textit{bliss} \cite{bliss, bliss_2}.
specifically, \textit{bliss} computes the canonical representative map function $\rho$, wherein the following two conditions are applicable:
\begin{itemize}
  \item the representative of a graph $\rho(\mathcal{G})$ is isomorphic to graph $\mathcal{G}$ and
  \item the representatives of two graphs, $\rho(\mathcal{G}_1)$ and $\rho(\mathcal{G}_2)$, are identical if and only if the graphs, $\mathcal{G}_1$ and $\mathcal{G}_2$, are isomorphic.
\end{itemize}
In \cite{bliss}, it is demonstrated that \textit{bliss} performs canonical labeling.

%
%

\section{Reinforcement Learning}
\label{sec:rl-problem}
In this section, we explain the outline of the reinforcement learning \cite{sutton}.
Reinforcement learning corresponds to a learning method to obtain a policy that determines the action for each time step.

In reinforcement learning problem, the state value function $V^{\pi}(s)$ of the policy $\pi$ is defined as the expectation of cumulative reward as follows:
\begin{align}
  V^{\pi}(s) = \mathbb{E}^{\pi}\!\left[\sum^{\infty}_{t=0} \gamma^t r_{t+1}\,\middle|\,s_0 = s\right],
\end{align}
where $\gamma \in \left[0,1\right]$ denotes a discount rate, which is a parameter that denotes how valuable the future rewards are at current state.
The action value function $Q^{\pi}(s,a)$ is defined as the value of taking action $a$ in state $s$ based on policy $\pi$ as follows:
\begin{align}
  Q^{\pi}(s,a) = \mathbb{E}^{\pi}\!\left[\sum^{\infty}_{t=0} \gamma^t r_{t+1}\,\middle|\,s_0 = s, a_0 = a\right].
\end{align}
When the following inequality that relates to two policies $\pi_1$ and $\pi_2$ is established, $\pi_1$ is considered as better than $\pi_2$,
\begin{align}
  \forall s \in \mathcal{S},\forall a \in \mathcal{A},\ Q^{\pi_1}(s, a) \geq Q^{\pi_2}(s, a).
\end{align}
The goal of reinforcement learning involves obtaining the optimal policy $\pi^*$ that maximizes $Q^{\pi}(s, a)$ as follows:
\begin{align}
  \pi^*(a\,|\,s) &= \argmax_{a\in\mathcal{A}} Q^*(s,a) \nonumber \\
  &= \argmax_{a\in\mathcal{A}} \mathbb{E}[\mathcal{R}(s,a,s') + \gamma V^{\pi^*}(s')].
\end{align}
Specifically, Q-learning is a reinforcement learning method to obtain an optimal policy $\pi^*(a\,|\,s)$.

\section{Proposed Scheme}
\label{sec:propose}
In the study, we use the DDQN as the deep reinforcement learning-based algorithm.
To extract the features of the graph structure of APs, we use GCNs \cite{gcn3} as layers of a neural network.
Additionally, we use SAP \cite{pg1,pg2} to collect training data to improve the learning speed.

\subsection{Algorithm}
\label{sec:alg}
In the study, we solve the problem defined in Section~\ref{sec:system_model} based on the deep reinforcement learning.
The baseline method corresponds to the deep Q-network (DQN) \cite{dqn}.

The main factors of the DQN include \textit{experience replay} and \textit{fixed target Q-network}.
Generally, Q-learning with function approximation may not converge \cite{experience_replay}.
Fixed target Q-network corresponds to a method that promotes convergence by fixing the target value used to calculate the error value, which should be minimized.
The DQN employs two networks, namely a main network $Q_{\bm{\theta}}$ (which is the target of the optimization with a set of weights $\bm{\theta}$) and a target network $Q_{\bm{\theta^{-}}}$ (which is used to calculate the temporal difference errors (TD errors) with a set of weights $\bm{\theta^{-}}$).
The parameter of the target network $\bm{\theta^{-}}$ is updated to $\bm{\theta}$ only every $I$ time steps and then maintained as fixed between updates.

Additionally, experience replay corresponds to a technique that breaks temporal correlation in the training data.
The training data ($s, a, r, s'$) is randomly sampled from the replay buffer $\mathcal{D}$, which stores the observed current state, action, reward, and next state ($s_t, a_t, r_{t+1}, s'_{t+1}$) at each time step $t$.
The interval $I$ of the target update and the size of the replay buffer $\mathcal{D}$ corresponds to user-defined parameters, which are detailed in Section~\ref{sec:evaluate}.

In the DQN, the parameter $\bm{\theta}$ is updated in each time step $t$ as follows:
\begin{align}
  \bm{\theta} \leftarrow \bm{\theta} + \alpha \left( Y_t^{\mathrm{Q}} - Q_{\bm{\theta}}(s_t, a_t) \right) \nabla_{\bm{\theta}}Q_{\bm{\theta}}(s_t, a_t), \\
  Y_t^\mathrm{Q} := r_{t+1} + \gamma \max_a Q_{\bm{\theta}} (s_{t+1}, a).
\end{align}

In addition to the original DQN, we employ several well-known techniques as follows:
\textit{DDQN} \cite{ddqn2}, \textit{dueling network} \cite{dn}, and \textit{prioritized experience replay} \cite{per},
which are known to contribute to the general performance improvement of DQN.
By using these techniques, we can avoid overestimations, learn the values of the states without the effect of actions, and sample the more effective training data to learn from the replay buffer $\mathcal{D}$.
Their details are described in the Appendix.

\subsection{Graph Convolutional Networks}
\label{sec:gcn}
In this section, we describe the function approximation method that is suitable for the state designed based on the adjacency matrix.

Specifically, GCN corresponds to the method that extracts the features of signals defined on the graph\cite{gcn3}.
In our system model, the topology of APs is expressed through a graph, and the number of the graphs available is extremely high.
Therefore, we analyze the graph structure based on a GCN, such as convolutional neural networks \cite{cnn}, which extracts the features of images.

Generally, the convolution calculation in time domain is expressed as the Hadamard product in the frequency domain.
Therefore, GCN is expressed by applying an inverse Fourier transformation to the result that corresponds to the Hadamard product after the Fourier transformation.
If the input dimension corresponds to $d \in \mathbb{R}$, the following process is adapted to each dimension.

An input vector $\bm{x} \in \mathbb{R}^N$ corresponds to a signal of the graph $\mathcal{G}$ with $N$ nodes.
Let $\bm{D}$ be an $N \times N$ degree matrix of the graph, and let $\bm{L} = \bm{D} - \bm{A}$ be its graph Laplacian with the adjacency matrix $\bm{A}$ of the graph $\mathcal{G}$.
Let the graph Laplacian $\bm{L}$ be orthogonally transformed as $\bm{L} = U^\mathsf{T}\bm{x}U$ with eigenvectors $U = (u_1, u_2, \dots, u_N)$.
Subsequently, a graph convolution of input signal $\bm{x}$ is defined as $\bm{x} \rightarrow U(\theta \odot (U^\mathsf{T} \bm{x}))$, where $\theta = (\theta_1, \dots, \theta_N)$ corresponds to the coefficient vector, and $\odot$ represents the Hadamard product.

By using GCN, the learning performance exceeds than that of a simple neural network, which consists of only fully connected layers (details are described in Section~\ref{sec:evaluate}).


\subsection{Data Collecting Policy}
\label{sec:sap}
Specifically, $\epsilon$-greedy corresponds to a general method to collect the training data \cite{sutton}.
Furthermore, $\epsilon$-greedy corresponds to the method that randomly selects an action with probability $\epsilon$ and selects the greedy action with probability $1-\epsilon$.
Given that Q-learning corresponds to an off-policy method, a degree of freedom exists in the method to collect the training data.
The employment of a better data collecting policy can make it possible to observe the better state that cannot be observed when the policy is $\epsilon$-greedy.

Thus, SAP corresponds to a potential game-based channel selection method that increases the throughput\cite{pg1,pg2}.
Given that SAP is independent of the learning network, it is expected to improve the learning speed via selecting appropriate actions even in an early stage of learning.
Based on potential game-based methods, each AP selects an action based on a best response to maximize its own throughput.
Therefore, it is possible that the agent reaches a non-optimal Nash equilibrium.
Conversely, SAP can prevent the agent from staying in a non-optimal Nash equilibrium by stochastically selecting the action.
In the study, the payoff function used in SAP is based on\cite{pg2}.
The probability that the AP\ $i$ selects channel $c_j$ is expressed as follows:
\begin{align}
  \label{eq:sap}
  &P(i, c_j) = \frac{\exp [\beta u_i (c_j, c_{-j})]}{\sum_{c'_j} \exp [\beta u_i (c'_j, c_{-j})]} , \\
  &u_i(\bm{c}_\mathcal{N}) := -\sum_{j \neq i} \ind(c_j = c_i) \ind(e_{ij} \in \mathcal{E}) , \\
  &\bm{c}_{-i} := ( c_1, \dots, c_{i-1}, c_{i+1}, \dots, c_N ) \nonumber ,
\end{align}
where $u_i(\bm{c}_\mathcal{N})$ denotes the payoff function; $\ind(x)$ denotes the indicator function that corresponds to one if event $x$ is true and corresponds to zero otherwise; $\bm{c}_i$ denotes the set of channels available to AP\ $i$; and $\beta \geq 0$ denotes the parameter that determines the degree of selecting the state with high payoff function.

\section{Simulation Evaluation}
\label{sec:evaluate}
\begin{figure}[tb]
  \centering
  \subfloat[Structure of a GCN-based model.]{
    \includegraphics[height=4.5cm]{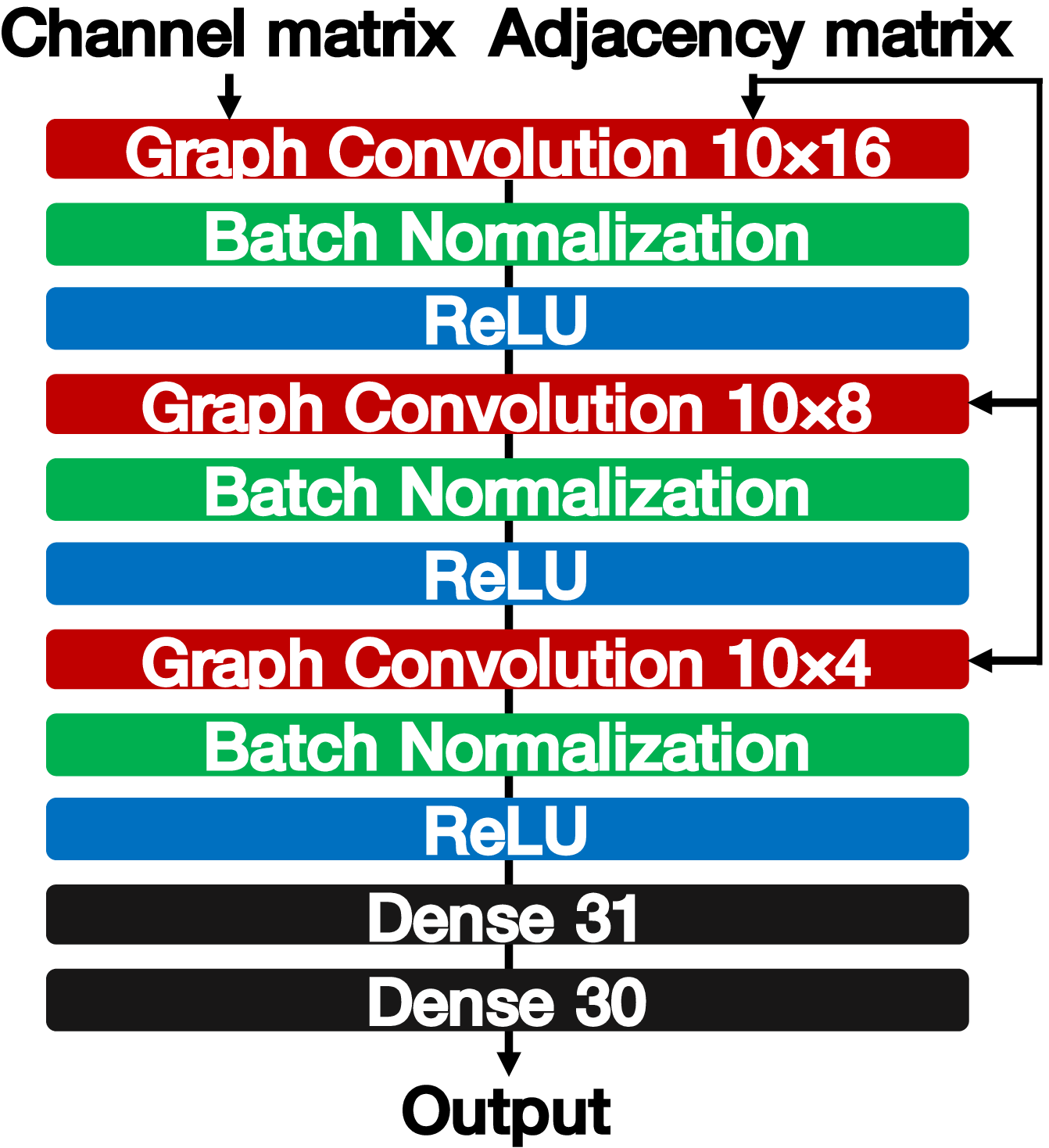}
    \label{fig:gcn_layer}
  }
  \centering
  \subfloat[Structure of a simple neural network model.]{
    \includegraphics[height=4.5cm]{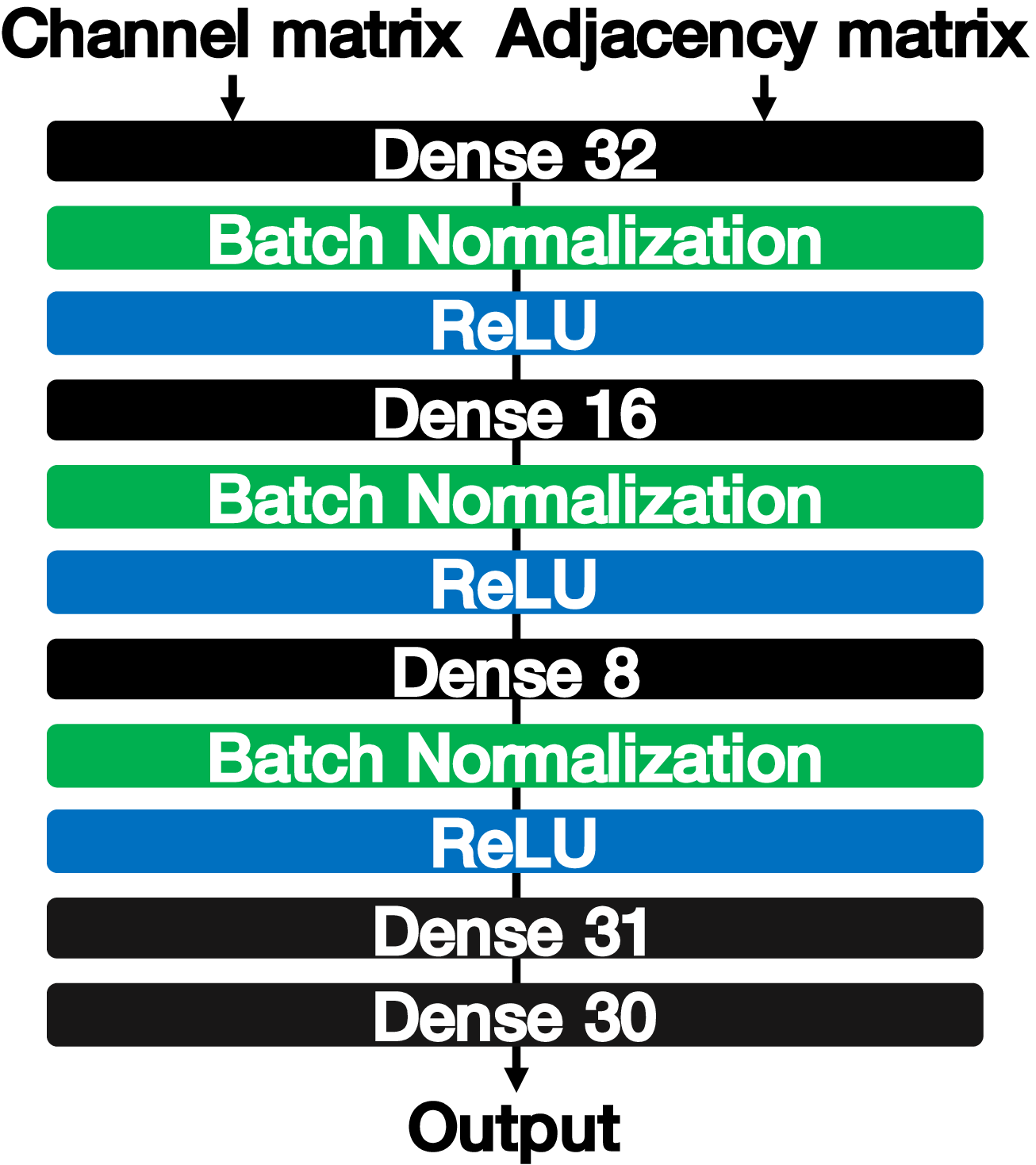}
    \label{fig:nn_layer}
  }
  \caption{Structures of GCN-based and simple neural network models. The input corresponds to the set of adjacency matrix $\bm{A}$ and channel matrix $\bm{C}$, and the outputs correspond to the estimated action values $Q(s,a)$ for each action $a$.}
  \label{fig:layer}
\end{figure}

In this section, we confirm the efficiency of the proposed scheme through proof-of-concept simulations.
Fig.~\ref{fig:layer} shows the overall architectures used in the simulations where Figs.~\ref{fig:layer}\subref{fig:gcn_layer} and \ref{fig:layer}\subref{fig:nn_layer} represent the GCN-based and simple neural network models, respectively.
Additionally, the simulation parameters are summarized in Table~\ref{tab:parameter}.
In the simulations, we evaluate the performance of the main network $Q_{\bm{\theta}}$ for every 10000 time steps, and complete learning if the maximum of the evaluated values is not updated for 300000 time steps.
The evaluated value corresponds to the mean value of 100 episode final rewards $R_\mathrm{m}$, which denote the rewards after 20-step greedy actions from the initial state.
\begin{table}[tb]
  \centering
  \caption{Simulation parameters}
  \begin{tabular}{ll}
    \toprule
    Parameter & Value \\
    \midrule
    Number of APs, $N$ & 10\\
    Number of available channels & 3 \\
    Carrier sensing range & 550\,m \\
    Scattering region of APs & 1000\,m $\times$ 1000\,m square-shape\\
    Throughput & BoE throughput \cite{BoE}\\
    Reward & Average throughput of the lower 4 APs\\
    Update period of $Q_{\bm{\theta^{-}}}$, $I$ & 100000 time steps\\
    Discount rate $\gamma$ & 0.9 \\
    $\beta$ of (\ref{eq:sap}) & 0.1 \\
    Batch size & 32 \\
    Optimizer & Adam \cite{adam} (learning rate = 0.001) \\
    Loss function & Huber loss \\
    $\epsilon$ of $\epsilon$-greedy & 0.1 \\
    Replay buffer size & 1000 \\
    \bottomrule
  \end{tabular}
  \label{tab:parameter}
\end{table}

We use the back-of-the-envelope (BoE) throughput \cite{BoE} as the throughput of APs.
The BoE throughput corresponds to a value that allows us to adopt shortcuts in performance evaluation and bypass complicated stochastic analysis.
We evaluate the target network $Q_{\bm{\theta^{-}}}$ after learning based on the reward, which corresponds to the reward at state after 20-step greedy actions from the initial state.
We use $\epsilon$-greedy \cite{sutton} as a comparison method for collecting training data.
We compared the results of two models in Fig.~\ref{fig:layer} and four methods as follows: the deep reinforcement learning-based method (which selects the action based on SAP and $\epsilon$-greedy); a simple SAP method according to potential game theory; and random action selection method.

\begin{figure}[tb]
  \centering
  \includegraphics[width=0.38\textwidth]{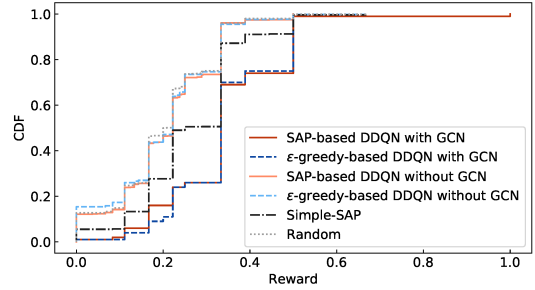}
  \caption{CDFs of the final rewards $r_{21}$ for 1000 episodes. The proportions of the high reward state of the results of the deep reinforcement learning-based methods with the GCN-based model exceed those of the other methods.}
  \label{fig:result}
\end{figure}

The cumulative distribution functions (CDFs) of the rewards for 1000 episodes are shown in Fig.~\ref{fig:result}.
As shown in the figure, the proportions of the high reward state of the results of the deep reinforcement learning-based methods with GCN-based model exceed those of the other methods.
By using the GCN-based model, the learning performance exceeds that of the simple neural network model.

\begin{figure}[tb]
  \centering
  \includegraphics[width=0.38\textwidth]{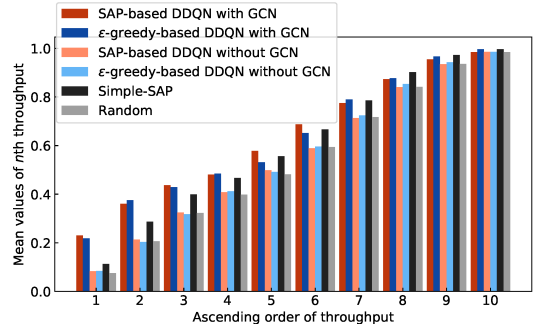}
  \caption{Mean values of $n$th lowest throughput.}
  \label{fig:all_rew}
\end{figure}

The mean values of the $n$th lowest throughput in all episodes are shown in Fig.~\ref{fig:all_rew}.
This figure shows that the mean values of $n$th lowest throughput of the deep reinforcement learning-based methods with GCN-based model are same as those of the other methods or larger.
Especially, the first to fourth lowest throughput increase by using the deep reinforcement learning-based methods with GCN-based model.
While focusing on the mean value of the lowest throughput, the GCN-based models achieve more than twice as the throughput of the other models.

\begin{figure}[tb]
  \centering
  \includegraphics[width=0.38\textwidth]{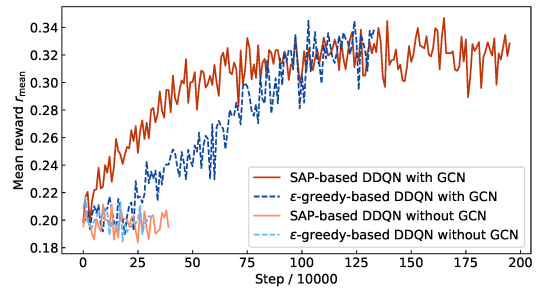}
  \caption{Transition of mean rewards $R_\mathrm{m}$ during learning. The increasing rate of $R_\mathrm{m}$ of the SAP-based method with the GCN-based model is the highest among the four DQN-based methods.}
  \label{fig:log}
\end{figure}

Fig.~\ref{fig:log} shows the learning curve that represents the transition of the mean reward $R_\mathrm{m}$ during learning.
The mean rewards $R_\mathrm{m}$ of the GCN-based models increase when the learning proceeds while that of the simple neural network models exhibit almost no increase.
While focusing on the GCN-based models, the increasing rate of $R_\mathrm{m}$ of the SAP-based method exceeds that of the $\epsilon$-greedy-based method.
This is potentially attributed to the fact that the SAP-based method does not refer to the action value function and might be able to select a better action even in an early stage of learning.
Given that the agent experiences more desirable states and actions to update the main network $Q_{\bm{\theta}}$, the overall rewards increase when compared to that of the $\epsilon$-greedy-based method and especially in an early stage of learning.

\section{Conclusion}
\label{sec:conclusion}
The study proposed a deep reinforcement learning-based channel allocation scheme in high density WLANs.
First, to extract the features of carrier sensing relationship, we applied GCNs to a contention graph in which APs within their carrier sensing ranges were connected.
Second, we used SAP to collect the training data to improve the learning speed.
Finally, we reduced the number of the states based on canonical labeling, which determined the isomorphism between graphs.
The simulation results indicated that our proposed scheme led to proportions of high reward states (which were states after 20-step greedy actions from the initial states) that exceeded those of the compared methods.
When focusing on the mean value of the lowest throughput, the GCN-based models achieve more than twice as the throughput of the other models.
By using GCN, we improved the learning performance when compared to that of the simple neural network model, which consists of only fully connected layers.
Additionally, we improved the performance more quickly by collecting the training data based on SAP while learning the model.

\appendix
\paragraph{DDQN}
In the study, we use the DDQN \cite{ddqn2} that corresponds to a DQN-based method to avoid overestimations.
The DDQN properly uses two networks: the main network $Q_{\bm{\theta}}$ to select actions with a set of weights $\bm{\theta}$, and the target network $Q_{\bm{\theta}^{-}}$ to evaluate the actions with a set of weights $\bm{\theta}^{-}$．
The error value of the DDQN is expressed as follows:
\begin{align}
  \label{eq:ddqn}
  Y_t^{\mathrm{DDQN}} &:= r_{t+1} + \gamma Q_{\bm{\theta^{-}}}(s_{t+1}, \argmax_a Q_{\bm{\theta}}(s_{t+1}, a)).
\end{align}

\paragraph{Dueling Network}
A dueling network \cite{dn} corresponds to a method that can learn which states are (or are not) valuable without entailing to learn the effect of each action for each state.
A dueling network includes two streams to separately estimate state-value and advantages for each action in neural network architecture.
The output value corresponds to the total value of the two streams.

\paragraph{Prioritized Experience Replay}
In the study, we sample the training data from the replay buffer $\mathcal{D}$ according to PER \cite{per}, which allocates priority to all samples based on the TD errors.
TD error $\delta_t$ in the DDQN is expressed as follows:
\begin{align}
  \delta_t = r_{t+1} + \gamma Q_{\bm{\theta^{-}}}(s_{t+1}, \argmax_a Q_{\bm{\theta}}(s_{t+1}, a)) - Q_{\bm{\theta^{-}}}(s_t, a_t).
\end{align}
The probability of sampling increases when the absolute value of TD error $|\delta_t|$ increases.
It should be noted that the probability of selecting the sample $i$ is expressed as follows:
\begin{align}
  P(i) = \frac{p_i^\lambda}{\sum_k p_k^\lambda} , \\
  p_i = |\delta_i| + \epsilon_0 ,
\end{align}
where $\epsilon_0$ denotes a tiny positive number that prevents the sampling probability from corresponding to zero when the TD error is zero.
Furthermore, $\lambda$ is the parameter that determines the degree of prioritizing for sampling, (for e.g., when $\lambda = 0$, the sampling is uniformly randomly implemented).

\section*{Acknowledgment}
This work was supported in part by JSPS KAKENHI Grant Numbers JP18H01442.
\bibliographystyle{IEEEtran}
\bibliography{vtcf_2019}

\end{document}